\documentclass{aa}  
\usepackage{graphicx}
\usepackage{lscape}
\usepackage[varg]{txfonts}

\usepackage{natbib} 
\bibpunct{(}{)}{;}{a}{}{,} 

\usepackage{ulem}

\begin{document}

\title{Dark matter in the Milky Way: Measurements
up to 3 kpc from the Galactic plane above the Sun}

\titlerunning{Mass Density of Dark Matter within the Milky Way}

\author{O. Bienaym\'e\inst{1}  \and A. C. Robin\inst{2} \and J.-B. Salomon\inst{3}  \and C. Reyl\'e\inst{2}  }

\institute{Observatoire astronomique de Strasbourg, Universit\'e de Strasbourg, CNRS,  11 rue de l'Universit\'e, F-67000 Strasbourg, France 
        \and
Institut Utinam, CNRS UMR 6213, Universit\'e de Franche-Comt\'e, OSU THETA Franche-Comt\'e-Bourgogne, Observatoire de Besan\c{c}on, BP 1615, 25010 Besan\c{c}on Cedex, France
    \and
Independent scholar    
 }

\date{}
\abstract
   {
We  probe the gravitational force perpendicular to the Galactic plane at the position of the Sun based on a sample of red giants, with measurements taken from the DR3 {\it Gaia} catalogue.  
Measurements far out of the Galactic plane up to 3.5 kpc allow us to determine directly the total mass density, where dark matter is dominant and the stellar and gas densities are very low.
In a complementary way, we have also used a new determination of the local baryonic mass density to help determine the density of dark matter in the Galactic plane at the solar position.
For the local mass density of dark matter, we obtained  $\rho_\mathrm{dm}$=0.0128$\pm $0.0008\,M$_\sun$\,pc$^{-3}$
= 0.486 $\pm$0.030 Gev\, cm$^{-3}$. For the flattening of the gravitational potential of the dark halo, it is $q_\mathrm{\phi,h}$=0.843$\pm0.035$. For its density, $q_\mathrm{\rho,h}$=0.781$\pm$0.055.}


\maketitle

\section{Introduction}

The determination of the gravitational potential perpendicular to the Galactic plane has an extensive history covering more than a century, beginning with  the work of \citet{kap22}. He noted the analogy between the Boltzmann equation 1D1V for the vertical motion of stars and the barometric equation for a plane-parallel atmosphere.  This work was immediately followed by a first determination of the force perpendicular to the Galactic plane, measured as $K_z$, by \citet{oor32}. Numerous studies have followed this thread and continued to this day.
A few noteworthy publications that have made it possible to trace many of these prior works include:
\citet{1986MNRAS.221.1023K}, \cite{1995IAUS..164..195K}, \cite{2000MNRAS.313..209H}, \citet{2014A&A...571A..92B}, 
\citet{2014JPhG...41f3101R}, \citet{2016ApJ...824..116K}, and \cite{2021RPPh...84j4901D}.

Several important steps have led to significant progress and improvements with respect to the measurement of $K_z$. The use of the same sample of tracer stars to simultaneously measure its density distribution and kinematics perpendicular to the Galactic plane by \citet{1989MNRAS.239..605K}  has done away with the uncertainties related to the lack of homogeneity of  combined samples from previous studies.

The development of a wide range of methods has improved the analysis of stellar samples. The distribution functions, which are the exact solutions of the 1D1V Boltzmann equation developed by \citet{kap22} and \citet{oor32} taking the form of stationary  (also known as isothermal) solutions, remain a satisfactory method that allows  for smoothing.
A separate approach is based on 1D Jeans equations \citep{2016MNRAS.458.3839X, 2016ApJ...817...13S}  which, in principle, avoids putting an a priori on the solutions and also avoids having to provide an explicit form of the distribution functions. In practice, due to the noise of  measurements, a smoothing of the data (density and kinematics as a function of distance from the Galactic plane) proves necessary. Thus, it should  be noted that the smoothed curves do not necessarily correspond to a stationary solution.

Another method is the direct inversion, which is undoubtedly very elegant, but which relies on an inverse Laplace transform that is known to be an ill conditioned numerical problem. In practice, just a small amount of  noise in the measurements of density and kinematics introduces large fluctuations in the recovered potential. Some examples of this problem can be found in the proceedings of a workshop on the  $K_z$ problem \citep[][see in particular the cover page of these proceedings]{1989gfpg.conf.....P}.

A global approach based on more complete modelling of the Galaxy, combining stellar counts and their kinematics, has also been used to measure the gravitational potential and in particular the force perpendicular to the Galactic plane; for example, in
\citet{1986A&A...157...71R}, \citet{1987A&A...180...94B}, and \citet{1989A&A...211....1C}. It has also become more commonly developed today \citep{ 2020MNRAS.494.6001N,   2022A&A...666A.130S, 2023MNRAS.520.1832B}.
Significant advances have  been made in the consideration of the coupling of radial and vertical motions \citep{1989MNRAS.239..605K}. At present, one of these approaches involves 3D modelling based on developing distribution functions dependent on three integrals of motion 
\citep{ 2014MNRAS.441.3284S, 2015A&A...581A.123B, 2022A&A...667A..98R, 2023MNRAS.520.1832B}.
These methods are necessary to fully describe the 3D velocities of stars and to accurately measure the potential beyond 500 pc outside the Galactic plane.

Nowadays, the \textit{Gaia} DR3 survey is several orders of magnitude more accurate than any other survey in terms of distances, 3D kinematics, definition of homogeneous samples, and number of stars measured. 
This survey has been used in recent publications  to develop new methods for analysing the $K_z$ forces.
A key independent observation was the measurement of $R_0$, the distance from the Sun to the Galaxy's central black hole 
\citep{2018A&A...615L..15G, 2019A&A...625L..10G}.  This has finally enabled the community to ascertain the shape of the rotation curve at different Galactic radii
\citep{ 2018RNAAS...2..156M, 2019ApJ...870L..10M,  2019ApJ...871..120E, 2021MNRAS.501.5964D}.
It is both the knowledge of the slope of the Galactic rotation curve and measurement of the $K_z$, the vertical force, that make it possible (via Poisson's equation) to determine the total mass density in the solar neighbourhood. 
An uncertainty in the slope of the rotation curve $V_c\sim R^\alpha$ equivalent to an uncertainty of about $\pm$0.01 in the $\alpha$ coefficient leads to an uncertainty of $\mp 0.003 M_\odot /$pc$^3$ on the local mass density; this is a significant fraction of the expected dark matter density.
This $\alpha$ coefficient was known to this low precision before the GRAVITY measurements.

The {\it Gaia} observations have outlined the non-stationarity of the stellar disc, which is a limitation to all the previously mentioned methods. 
The identification of a spiral \citep{2018Natur.561..360A}  in the $z$-$w$ position-velocity space highlights the velocity fluctuations in the Galactic plane. \citet{2012ApJ...750L..41W} had already noted asymmetries in counts between the directions of the north and south Galactic poles as tracers of vertical waves in the Galactic disc.  However, this has opened up a new approach using the properties of spirals in $z$-$w$ space to measure $K_z$ independently of other methods \citep{2019MNRAS.482..262W,2021A&A...653A..86W,2022A&A...663A..16W,2024ApJ...960..133G}.

All of these approaches to measuring the gravitational potential in the disc only determine the total mass density, baryonic and dark matter, without distinguishing between them.  As long as we remain close to the Galactic plane, most of the mass is baryonic (stellar and ISM). At $R_0$ and $z$=0 (in Galactic coordinates), we would expect only about 10 per cent of the mass density to be dark matter, assuming a spherical dark halo to explain the Galactic rotation curve.
 However, the uncertainty on the local stellar and ISM volume mass density was  known with a precision  of the order of the amount of dark matter. To our knowledge, since the work of \citet{2006MNRAS.372.1149F}  as well as \citet{2015ApJ...814...13M}, it is now only the work based on the local and distant stellar counts {\it Gaia} DR3 carried out by \citet{2022A&A...667A..98R} that has provided a more realistic estimate of the local stellar volume and surface mass density, which we refer to later in this paper. 

In this work, we propose an improvement to the measurement of dark matter density based on a determination of the gravity field at large distances from the Galactic plane up to 3.5 kpc, where dark matter is dominant. At these distances above the Galactic plane, the contribution of the baryonic mass becomes negligible and non-stationary effects are also expected to be weaker for stellar populations that have large velocity dispersions.
The very high precision of {\it Gaia} observations considerably simplifies the measurement of the $K_z$. It is therefore all the more necessary to take proper account of the systematic biases identified \citep{ 2021A&A...649A...4L, 2021A&A...654A..20G, 2023A&A...680A.105K}. These biases are far from negligible for parallaxes at distances of several kpc. Here, we apply the various corrections proposed by \citet{ 2021A&A...649A...4L}, which depend on the magnitudes, colours, and equatorial coordinates. 

Section 2 details the samples used with  the density and kinematic profiles measured. Section 3 presents the dynamical model for measuring the potential, with the method being a reworking of that developed in \citet{2015A&A...581A.123B,2018A&A...620A.103B} and \citet{,2022A&A...667A..98R}.
 A discussion on the baryonic mass distribution is given in Section 4. A general discussion and our conclusion are given in Sections 5 and 6.
\section{Data sampling}  
In this work, the measurement of the gravitational potential and the vertical force $K_z$ is based  on the main assumption that the distribution of stars is in a stationary state. A sample that is homogeneous in magnitude and colour is requisite for this measurement, noting that homogeneous refers to the statistical point of view, namely, that the sample would be drawn from a stationary distribution law of a model of the Galaxy. This results in  a sample of stars that all have the same properties in terms of colour and absolute magnitude, while the outline of the spatial domain under analysis is precisely defined.
The {\it Gaia} DR3 catalogue makes it possible to constitute such a sample with an accuracy that has not been  achieved thus far for this type of study, thanks to the high accuracy of its measurements.

\begin{figure}[htbp]
   \centering
  \includegraphics[scale=0.35]{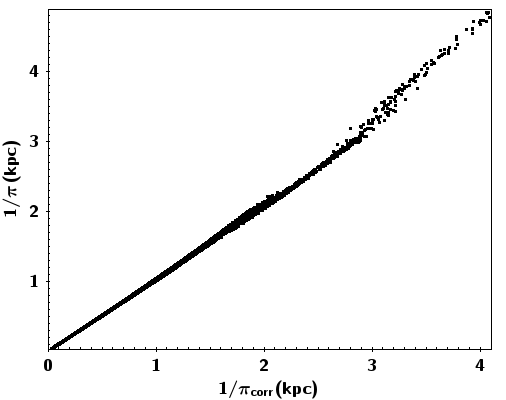} 
   \caption{ Stellar samples towards the Galactic poles: distances from {\it Gaia} DR3 parallaxes versus distances determined with corrected parallaxes (correction of 7\% at 3 kpc and 20 \% at 4 kpc).}
   \label{fig:f1}
\end{figure}

To probe  distances with sufficiently large sample sizes, we select the giant stars of the 'red clump'. The only significant source of error in this study concerns a bias in the DR3 {\it Gaia} parallaxes for the most distant stars. This bias has been analysed by \citet{2021A&A...649A...4L}  and \citet{2021A&A...654A..20G}, among others, revealing systematic errors that depend on the colour, magnitude, and position of the stars. This translates into a relative systematic error that increases with the distance of the stars. For stars located at 4\,kpc, this error on the distance can vary from 0 to 40 micro-arcsec (up to 20 per cent on distances at 4\,kpc towards the Galactic poles) depending on the stars. We have applied the parallax corrections (Figure~\ref{fig:f1}) proposed by \citet{2021A&A...649A...4L}.
They stress that a similar study should be undertaken for proper motions, for which there must also be a bias.
We  can admit that this bias must be  (at most) of the same order of magnitude as for the parallaxes, which translates into biases of less than 1\,km\,s$^{-1}$ at a distance of 4\,kpc. There, we consider this bias negligible.
The sample is defined by the colour-magnitude interval that covers the red
clump giants. We have chosen the infrared magnitudes $J$ and $K$ from the
2MASS survey, which minimises the effects of absorption by the interstellar
medium; towards the poles, the absorption in the $K$ band is of the order
or smaller than 0.012 \citep{2008A&A...488..935G}. Thus, it is neglected.\\

\begin{figure}[htbp]
   \centering
   \includegraphics[scale=0.35]{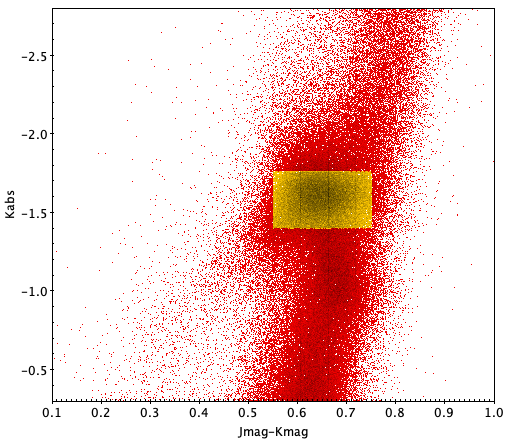} 
   \caption{  $J$ an $K$ magnitudes for stars with $|z|<$4 kpc and $|b|>$30 degrees. The sample selected by the magnitude limits is coloured yellow.}
   \label{fig:figRef1}
\end{figure}

\begin{figure}[htbp]
   \centering
   \includegraphics[scale=0.24]{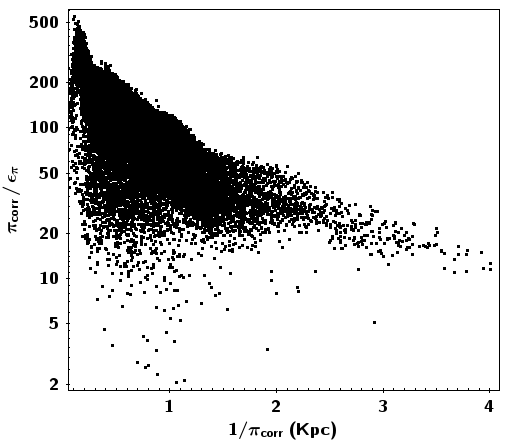} 
    \includegraphics[scale=0.24]{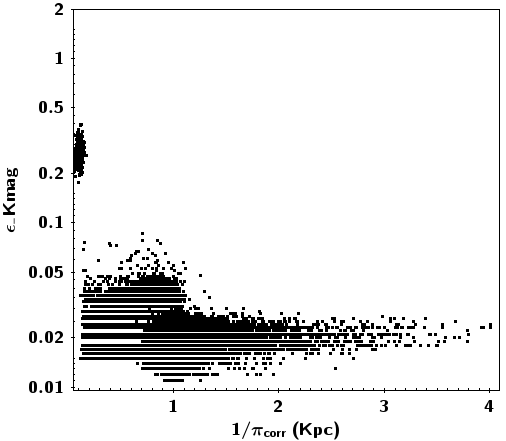} 
   \caption{ Inverse of the relative error on parallaxes, $\pi/\epsilon_\pi$, versus distances (left).
    Uncertainties on apparent 2MASS $\mathrm{K}$ magnitudes (right).
   }
   \label{fig:f2}
\end{figure}

\begin{figure}[htbp]
   \centering
   \includegraphics[scale=0.24]{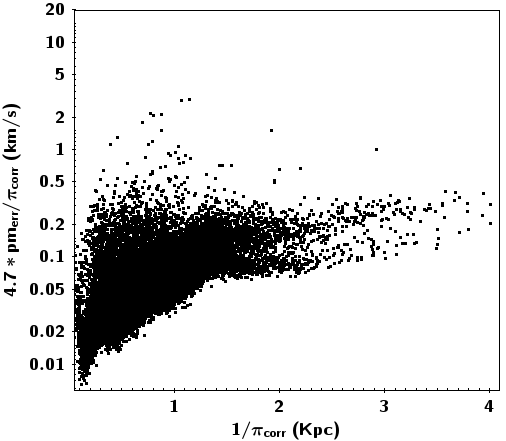} 
   \includegraphics[scale=0.24]{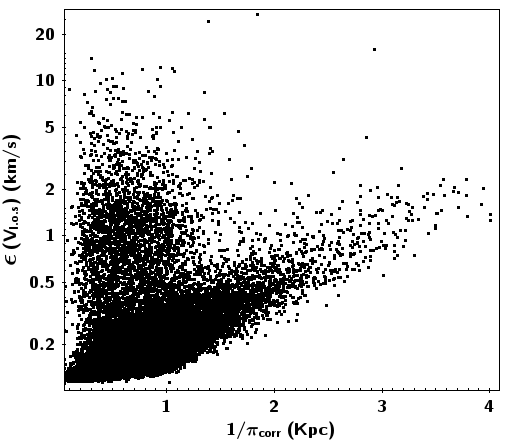} 
   \caption{ Uncertainties on transverse velocities from proper motions (left) and on the radial velocities versus distances (right).
   }
   \label{fig:f3}
\end{figure}

We set the absolute magnitude to
$K_{abs} = K_{mag} +5 * \log_{10}(\pi_{corr})-10 $ \, ,
where $\pi_{corr}$ is the parallax of {\it Gaia} DR3 corrected according to the recommendations of \citet{2021A&A...649A...4L}.
The selected samples consist of stars of colour $J-K$ ranging from 0.55 to 0.75 and magnitude $K_{abs}$ from $-1.4$ to $-1.76$
(see Figure~\ref{fig:figRef1}).
Thanks to the precision of colour and distance (Figure~\ref{fig:f2} and\,\ref{fig:f3}), with relative errors of just a few percent, many sources of selection bias have been considerably reduced. Examples include the Malmquist or Lutz-Kelker bias, which are linked to sample limits.

The nearly completeness of our sample can be claimed, 
since our first selection from {\it Gaia}-DR3 in magnitude and colour ensure completeness for these directions and far away from the Galactic plane where the stellar density is low with nearly no overlap of stellar images. Then considering the DR3 sample  with |{\it b}|>22 degrees, {\it bp\_rp} within [1,1.5], and {\it phot\_g\_mean\_mag} $<$14, this sample covers our red clump sample and only  0.37\% of these stars have no parallax or radial velocity. Only a tiny fraction ($10^{-4}$) of the selected sources have no photometric counterpart in the 2MASS catalogue, which is itself complete in these unconfused regions \citep{2006AJ....131.1163S}. Thus,  our sample of red clump stars is nearly complete.

Here, we use  the Galactic Cartesian coordinates $(x, y, z ; U, V, W)$ computed from {\it Topcat} functions \citep{2005ASPC..347...29T} and the Galactocentric cylindrical coordinates $(R, \phi, z; V_R, V_\phi, V_z)$.
The position of the Sun is assumed to be $R_0$=8.1 kpc and $z_0$= 19 pc.
The velocity of the Sun relative to the local standard of rest (LSR) is chosen to be $U_\odot$=12.9 km\,s$^{-1}$, $V_\odot$=12 km\,s$^{-1}$, $W_\odot$=7.8 km\,s$^{-1}$, values which are obtained a posteriori after fitting the velocity distributions of our samples (see Section \ref{methods}).
The circular rotation velocity at $R_0$ is assumed to be $V_c(R_0)$=236.27 km\,s$^{-1}$, value adopted in \citet{2022A&A...667A..98R}   \citep[see also][]{2020MNRAS.494.6001N}.
\\

\begin{figure}[htbp]
   \centering
   \includegraphics[scale=0.24]{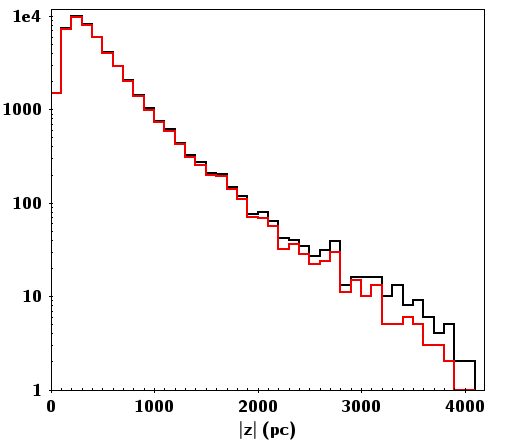} 
      \includegraphics[scale=0.24]{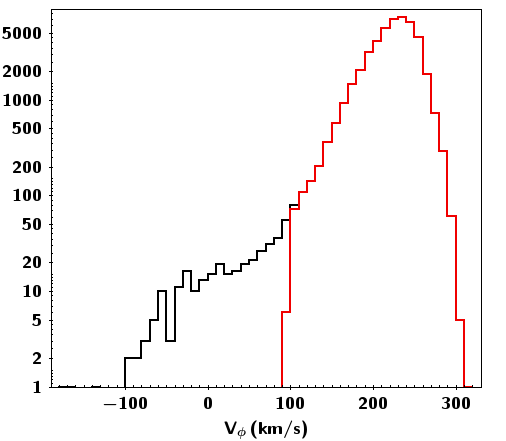} 
   \caption{Red clump stars towards the Galactic poles:  
   Histogram of star counts versus $|z|$ (left) and versus $V\phi$ (right). Red line: Stars with the highest $L_z>100\,\times\,8100$ km\,s$^{-1}$\,pc. Black line: All $L_z$.}
   \label{fig:f4}
\end{figure}

Two  samples towards the North and South Galactic poles (NGP, SGP) were used to measure the $K_z$ force and are obtained by
a selection on the Galactic latitude $|b|>22 \deg$, $|R_{Gal}-8.1| < 0.4$ kpc, $ |\phi| < 0.1234 $, (so $R_0\, \times \,\phi$ = 1\,kpc), and $ |z| <$ 4 kpc.
In addition, only the stars with the largest angular momentum are retained,
$L_z > 8.1$\,kpc $\times$ 100 km\,s$^{-1}$, to eliminate most of the halo stars (Figure~\ref{fig:f4}). We note that this last selection is  based on an integral of motion.

\begin{figure*}[htbp]
   \centering
   \includegraphics[scale=0.24]{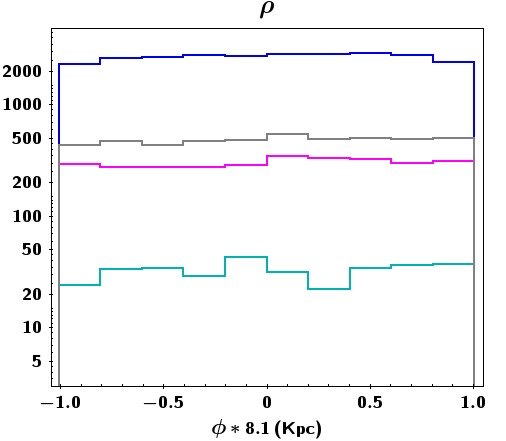} 
   \includegraphics[scale=0.24]{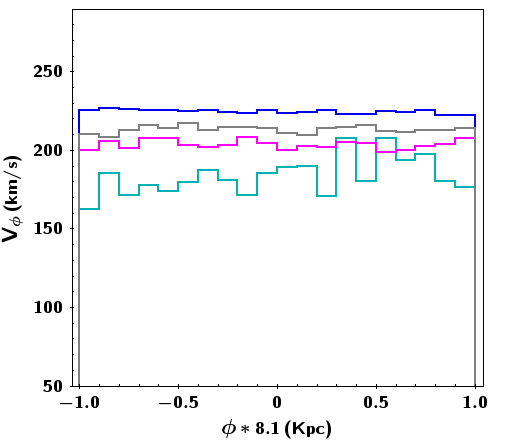} 
   \includegraphics[scale=0.24]{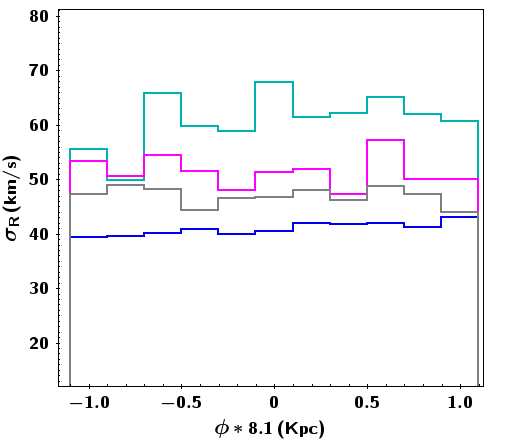} 
   \includegraphics[scale=0.24]{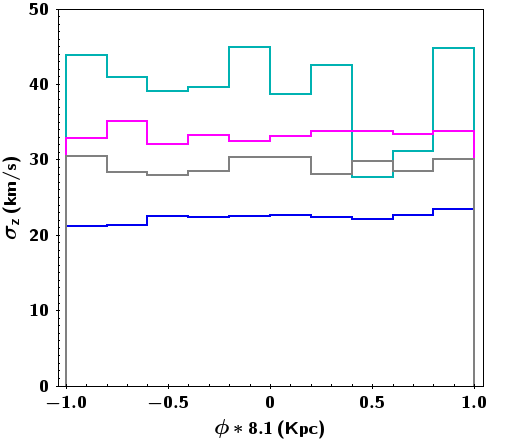} 
   \caption{ Red clump star properties towards the Galactic poles (NGP+SGP): from top left to bottom right, distribution of $\rho$, $ <V\phi>$, $\sigma_R$, and $\sigma_z$ with respect to $\phi$ at four values of $|z|$ (0.5, 1, 1.5 and 2.5 kpc) (resp. blue, grey, pink, and cyan lines).}
   \label{fig:f5}
\end{figure*}

For these samples, Figure~\ref{fig:f5} shows at different heights $|z|$, ($z$=500, 1000, 1500, 2500 pc), as a function of $\phi$, the density, $\rho$, the velocity dispersions,  $\sigma_R$, $\sigma_z$, and  $ <V\phi>$.
 This illustrates the relative constancy of these quantities, an indication of the degree of stationarity.
 
\begin{figure}[htbp]
   \centering
   \includegraphics[scale=0.24]{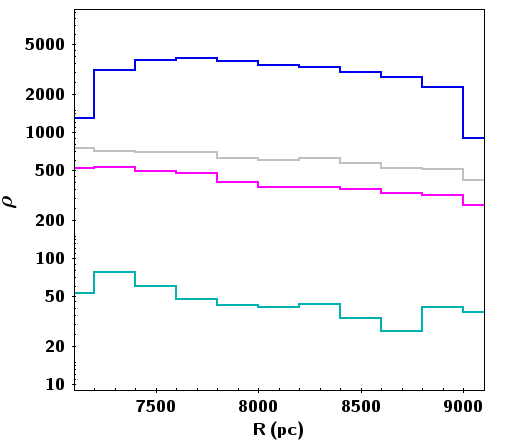} 
   \includegraphics[scale=0.24]{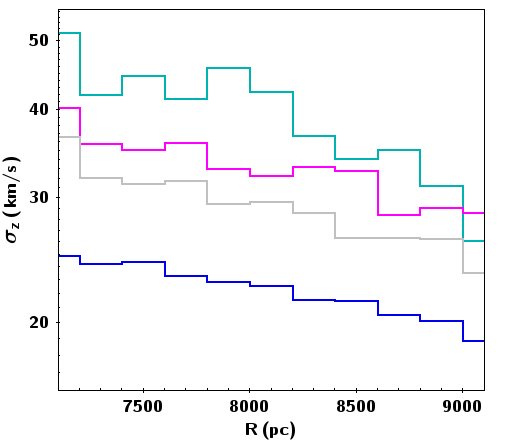} 
   \caption{Sample distribution (NGP+SGP) of $\rho$ and $\sigma_z$ along the Galactic radius. Same colour coding as in Figure~\ref{fig:f5}.}
   \label{fig:f6}
\end{figure}

A third sample was chosen according to the same selection in colour, magnitude, and $L_z$, but along the Galactic radius from 4 kpc to 12 kpc and an azimuth extension of $\pm500$\, pc. This sample is used to estimate the radial density and kinematic gradients at different $z$ heights (Figure~\ref{fig:f6}).

\section{Methods and model}   \label{methods}

\subsection{Distribution function}

To measure the vertical potential at the solar Galactic position up to a vertical distance of 3.5\,kpc, we  adapted and modified the century-old method developed by \citet{kap22} and \citet{oor32}. Indeed, their 1D1V method can only be applied when the stellar oscillations through the Galactic plane are smaller than $\sim1$\,kpc, where the vertical motions remain approximately decoupled from the horizontal ones.
By building exact 3D3V stationary solutions with Stäckel potentials, \citet{sta89} found corrections of the order of 10\% at 1 kpc compared with a 1D model.

Thus, for a correct modelling at larger $z$, we  developed  stellar population distribution functions (DFs)  for an axisymmetric gravitational potential which are 3D generalisations of  \citet{shu69} DFs. These DFs are inspired by \citet{sta89}. Our DFs of positions and velocities of each stellar disc are stationary and are modelled  with  three isolating
integrals  of motion:
\begin{equation} 
\label{eq:DF}
    \begin{tabular}{llll}
$f(E,L_z,I_3)= g(L_z)  \, \tilde\rho_{0} \, \exp\left({\frac{R_c-R_0} {\tilde{H}_{\rho}}}\right) \,  \exp\left({-\frac{E_{\parallel}}{\tilde{\sigma}_{R}^2}}\right)  \,  \exp\left({-\frac{E_{\perp}}{\tilde{\sigma}_{z}^2}}\right), $
 \end{tabular}
 \end{equation}
\begin{equation*}
    \begin{tabular}{llll}
with &
  $\tilde{\sigma}_R = \tilde{\sigma}_{0,R}  \exp\left({-\frac{R_c-R_0}{\tilde{H}_{\sigma_R}}}\right)$ ,\\
   &
    $\tilde{\sigma}_z = \tilde{\sigma}_{0,z}  \exp\left({-\frac{R_c-R_0}{\tilde{H}_{\sigma_z}}}\right)$ , \\
  and &
   $g(L_z) = \frac{\Omega }{\kappa } \frac{1}{2 \pi \, {\sigma_R}^2}$ \, .
\end{tabular}
\end{equation*}
\\
Here, $E_{\parallel} = (E-E_c)(I_{3,max}-I_3)$ and $E_{\perp} =  (E-E_c)I_3$ are integrals of motion depending on the total energy, $E$, as well as on $E_c$, the energy of the circular orbit of angular momentum, $L_z$, and  on $I_3$, which is an approximate St\"ackel integral \citep[details in ][]{2014A&A...571A..92B, 2015A&A...581A.123B,2018A&A...620A.103B}. Also,
 $E_{\parallel}$ and $E_{\perp}$  are linked respectively to   the radial and vertical motion of the stars;
$R_0$ is the Galactic radius at solar position and  $R_c(L_z)$ is the radius of the circular orbit with angular momentum, $L_z$. Then,
$\Omega$ and  $\kappa$ are the circular and epicyclic frequencies expressed as being dependent on $R_c(L_z)$ 
\citep[see Equation 8 in ][]{2015A&A...581A.123B}.
With St\"ackel potentials, $I_{3,max}$=1 corresponds to the shell orbit and $E=E_c+E_{\parallel} + E_{\perp}$. For non St\"ackel-potential such as the Galactic potential, $I_{3,max}(Lz)$ remains close to 1 and must be computed.
The distribution function $f(E,L_z,I_3)$ allows us to compute its various moments, which give us the density, $\rho(R,z)$, the rotational velocity, $V_{\phi}$, the velocity dispersions, $\sigma_R$, $\sigma_\phi$, and $\sigma_z$, and the tilt angle of the velocity ellipsoid. 
 For small velocity dispersions, the  density, and the radial and vertical velocity dispersions have a radial exponential decrease.
The three first input parameters, $ \tilde{\rho}_0$, $ \tilde{\sigma}_{R,0}$, and $ \tilde{\sigma}_{z,0}$, are  directly related (but not exactly equal) to the computed moments of the DFs at the solar position, respectively, the density $\rho(R_0,z=0)$ and the dispersions $\sigma_R(R_0,z=0)$ and $\sigma_z(R_0,z=0)$.
The other free parameters, $\tilde{H}_{\rho}$, $\tilde{H}_{\sigma_R}$, and $\tilde{H}_{\sigma_z}$,  of the stellar discs are related to the radial scale lengths for the density and for the kinematics. 

We note that the DF (Equation \ref{eq:DF}) is approximately  isothermal, but exactly isothermal in the case of a separable potential in $R$ and $z$ coordinates. It is also similar to the DFs used by \citet{2011MNRAS.413.1889B}   in the case of small departures from circular motions  where $E_\parallel \simeq  \kappa  J_R$ and $E_\perp  \simeq  \nu J_z$ , with $\kappa$ and $\nu$ the epicyclic and vertical frequencies, $J_R$ and $J_z$ the radial and vertical actions.

Finally,  we introduce a cut  and set to zero the DF (Equation \ref{eq:DF}) for orbits with    angular momentum larger than $L_{z,cut}$=8.1 kpc $\times$ 100 km\,s$^{-1}$. Actually, if this  DF is effective  to model  rotating flat stellar discs, it cannot be used to model the stellar halo; neither can it model  stars with small angular momentum. Here, we do not try to model the stellar halo with an ad hoc DF. Applying a cut in $L_z$, that is an integral of motion, allowed us to keep the stationarity of the DF, whereas we did not  model stars with small rotational velocities.

\subsection{Gravitational potential} 


The Galactic potential is involved in the distribution functions through the total energy, $E$ and $I_3$. The contributions to the potential come from the baryonic components, stars, and ISM, and the dark matter.  Here, we only constrain the  mass density of the dark matter component.

We consider that the stellar component is already well determined by the precise analysis of stellar counts in the immediate solar neighbourhood and at greater distances, obtained recently by fitting the Besan\c{c}on  Model of the Galaxy \citep[hereafter BGM,][]{2022A&A...667A..98R}.
On the other hand, the ISM component is less well characterised. We  adopt the stellar and ISM contribution to the gravitational potential as that obtained in \citet{2022A&A...667A..98R}. 

We model the potential of the dark matter halo in such a way that the Galactic circular rotation curve $V_c(R)$ also remains identical to that of the Besan\c{c}on model of the Galaxy. We set as a free parameter only its flattening, which makes it possible to adjust its local mass density, $\rho_\mathrm{dm}(R_0,z=0),$ without modifying the circular rotation curve. The analytical form of the dark matter halo is therefore slightly different from that of 
\citet[][see their Equation 4]{2022A&A...667A..98R}, which comes as follows:
$$ \Phi = -930300.8952 \, (R_\mathrm{dm}^2 + R^2 + z^2 / q_{\phi,h}^2 )^{-\gamma}, $$
with the parameter $q_{\phi,h}$ for the flattening of the dark halo. Then, 
$R_\mathrm{dm}$=3315 pc, $R,$ and $z$ are expressed in pc and $\Phi$ in (km.s$^{-1}$)$^2$.
The exponent $\gamma$=0.05 is used to fit the decreasing rotation curve of $R$ from 20 to 50 kpc \citep{2021MNRAS.501.5964D,2022A&A...667A..98R}.
\\ 
We note that if our fitting procedure modifies the density distribution inside the stellar disc, this only concerns stars
in the colour-magnitude range of the red clump giants.
It would only marginally modify the distribution of the total mass density in the Galactic disc that has been constrained and measured by \cite{2022A&A...667A..98R} and used here to determine the gravitational potential of the Galactic baryonic component.

\subsection{Adjustment of density and velocity profiles}

\begin{figure}[htbp]
   \centering
   \includegraphics[scale=0.24]{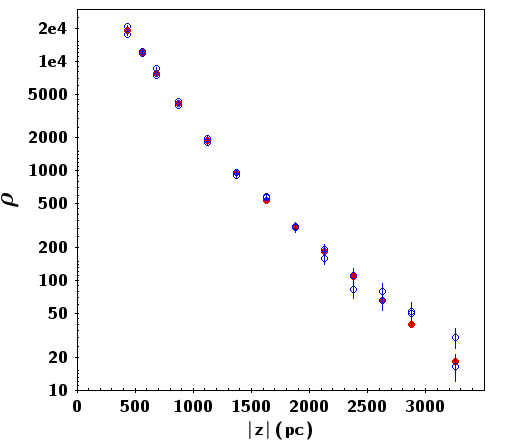} 
   \includegraphics[scale=0.24]{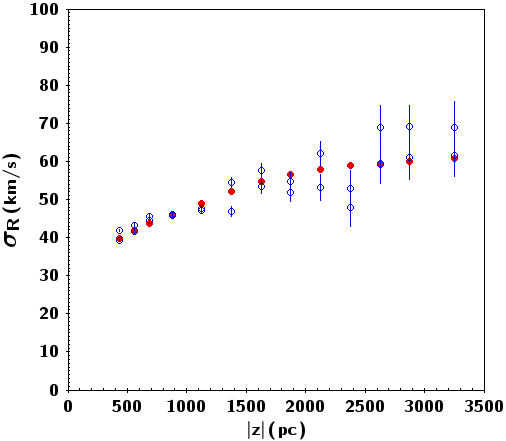} 
   \includegraphics[scale=0.24]{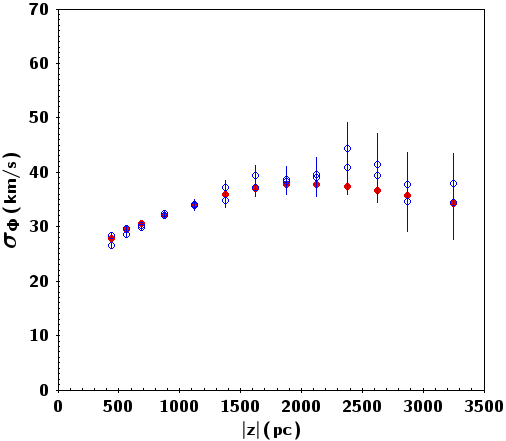} 
   \includegraphics[scale=0.24]{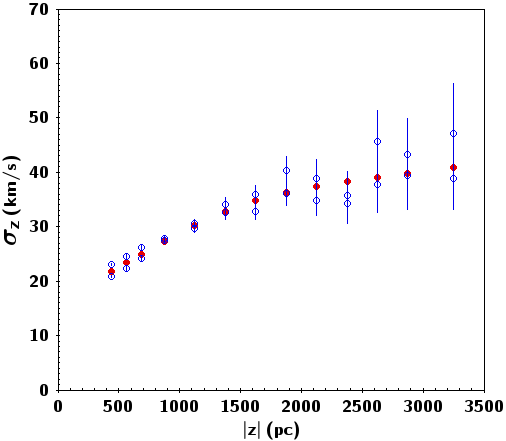} 
    \includegraphics[scale=0.24]{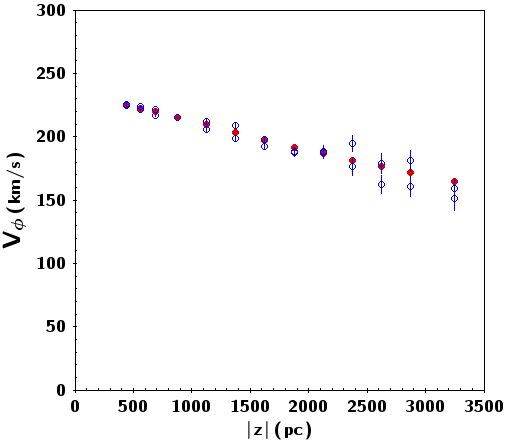} 
   \caption{Moments of the DFs of the NGP and SGP samples (two blue open circles at each distance) and  model  (red dot) for the  density, $\rho$, the velocity dispersions, $\sigma_R$, $\sigma_\Phi$,  
   $\sigma_z$, and the mean velocity, $V_\phi$.  }
   \label{fig:f7}
\end{figure}

 The distribution of the observed moments (density, mean Galactic rotational velocity, and the three velocity dispersions) towards the NGP and the SGP as a function of the distance to the Galactic plane, $z$, are fitted by summing four
elementary distribution functions given by Equation~\ref{eq:DF}, each corresponding to stellar discs of different thicknesses (Figure~\ref{fig:f7}). 
For each of the four elementary DFs, the fitted parameters are the two local velocity dispersions  at the solar position, 
$\tilde{\sigma}_{0,R}$ and $\tilde{\sigma}_{0,z}$,
 and its local density, $\tilde{\rho}_{0}$.
The fixed parameters are the radial scale lengths for the velocity dispersions and for the density. These fixed parameters are determined by adjusting the corresponding star counts within the Galactic meridian plane (two examples are shown in Figure\,\ref{fig:f6}).
The Sun velocity relative to the LSR is also a model parameter. It is fitted separately as detailed  in a following paragraph.

We determined the minimum $\chi^2$  for the parameters of the
model using the MINUIT software \citep{James2004}, which
 allows us to look for possible multiple minima and to obtain a determination of the variance-covariance matrix:
\begin{equation}
\begin{split}
    \chi^2 = \sum_i \, \biggl( \frac{(\rho_{mod,i}-\rho_{obs,i})^2} {\epsilon_{1,i}^2}
+\frac{(V_{\phi,mod,i}-V_{\phi,obs,i})^2}{\epsilon_{2,i}^2} \\
 +\frac{(\sigma_{R,mod,i}-\sigma_{R,obs,i})^2}{\epsilon_{3,i}^2} 
+\frac{(\sigma_{\phi,mod,i}-\sigma_{\phi,obs,i})^2}{\epsilon_{4,i}^2}\\
+\frac{(\sigma_{z,mod,i}-\sigma_{z,obs,i})^2}{\epsilon_{5,i}^2}
\biggr)
\end{split}
,\end{equation}
with $\rho_{obs,i}=N_i$, the number of stars in the $i$-th bin. The uncertainties are for density $\epsilon_{1,i} = \sqrt{N_i}$; for $V_\phi$, 
$\epsilon_{2,i} =\sigma_\phi /\sqrt{N_i}$;  for the dispersions $\sigma_{R(j=3)}$, $\sigma_{\phi(j=4)}$ and $\sigma_{z(j=5)}$, $\epsilon_{j,i} = \sigma_j/\sqrt{2N_i}$.
For each disc, the fitted parameters are related to the values at $z$=0 and $R_0$=8.1\,kpc of the density and the radial and vertical velocity dispersions, $\sigma_R$ and $\sigma_z$. The flattening $q_{\phi,h}$ of the dark halo is also adjusted.
The adjustments are performed by least squares method using both the north and south samples simultaneously. We note that the differences between these two distributions are greater than the uncertainty bars of each of these distributions.
The steps of the counting intervals are 125\,pc (for $z$ ranging from 375\,pc to 750\,pc), 250\,pc (from 750\,pc to 3000\,pc), and 500\,pc (from 3000\,pc to 3500\,pc).

The fit over the full interval from 375\,pc to 3.5\,kpc gives the flattening value $q_{\phi,h}=0.843\pm0.009$.
The other fitted or fixed parameters are given in Table \ref{default2}.
 Very small deviations from the fit in density can appear beyond $z$=2.5\,kpc (Figure~\ref{fig:f7}), which could indicate that the vertical dark matter distribution used (Equation \ref{eq:DF}) does not have sufficient degrees of freedom.
Otherwise, with the exception of the slight shift at large $z$ of the $\sigma_\phi$ distribution (see Figure~\ref{fig:f7}), the fits are very consistent with the data, lying mostly between the north and south distributions, and well within the error bar limits. This validates the distribution functions used. We obtain for the dark matter density
$\rho(z=0)_\mathrm{dm} $=0.0146\,M$_\odot$pc$^{-3}$.

\begin{table*}[htp]
\caption{  Best fitting values for stellar disc components. The remaining fitted parameters are the flattening of the dark halo potential, $q_\phi$, and the solar velocity (see text).}
\begin{center}
\begin{tabular}{ c | c c c | c c c  }
\hline\hline
& & Adjusted   &&& Fixed  & \\
Component & &  parameters  &&&  parameters & \\
 &  $\tilde\rho_{0}$  &  $\tilde{\sigma}_{R,0}$  &  $ \tilde{\sigma}_{z,0}$     &        $\tilde{H}_{\rho}$ & $\tilde{H}_{\sigma_R}$ &   $\tilde{H}_{\sigma_z}$ \\
   &  &  km/s &  km/s & kpc  & kpc  & kpc \\
\hline
1       & 3070$\pm$31   &               51.4$\pm$0.2                    & 28.7$\pm$0.1            & 2.9           & 21.5          & 6.1  \\
2       & 17119$\pm$112 &                       33.6$\pm$0.8            & 16.6$\pm$0.4            & 2.9           & 21.5          & 6.1  \\
3       & 21435$\pm$272 &                       32.3$\pm$0.2            & 10.8$\pm$0.1            & 4.2           & 62            & 5.8  \\
4       & 23635$\pm$808 &                        6.3$\pm$0.2                    & 7.1$\pm$0.08            & 2.5           & 70            & 3.5  \\
\hline
\end{tabular}
\end{center}
\label{default2}
\end{table*}%


 We also carried out a fit restricted to the interval 375\,pc - 2.5\,kpc for which we obtained a very close value, $q_{\phi,h}=0.839\pm0.016$, but with a higher uncertainty due to the smaller $z$ interval covered. The formal uncertainty, given by the variance-covariance matrix, is very low insofar as the parameter $q_{\phi,h}$ is strongly decorrelated from all the other parameters, with the exception of the dispersion $\sigma_z$ of the thickest disc.
 \\

We examine whether this result is sensitive to changes in the model.
The first set of fixed model parameters,  the  scale lengths,  are measured using the third stellar sample  along the Galactic radius. The radial gradient of $\sigma_R$ is  small and the uncertainty in the kinematic length scale, $H_{\sigma_R}$ does not change the adjusted parameters. 
On the other hand, varying the values of the $H_{\rho}$ and $H_{\sigma_z}$ scales within their uncertainties
affects the parameters related to the stellar discs, but not the determination of the $q_{\phi,h}$ parameter.

By simultaneously decreasing $H_{\rho}$ and $H_{\sigma_z}$, we can increase the contribution of stars with higher $\sigma_z$ dispersions coming from the inner parts ($R<R_0$). In this case, it has an impact on the measurement of the potential, but it is only significant for extreme changes in $H_{\sigma_z}$ and $H_{\rho}$ (beyond the uncertainties on these parameters).
The largest effect was obtained by increasing the scale length, $H_\rho$, of the thickest disc from 2.7\,kpc to 3.7\,kpc. The measurement of $q_{\phi,h}$  then rises from 0.84 to 0.85. We therefore consider that this uncertainty indicates that the systematic errors are less than 1\% on the measurement of the halo flattening and density.
Another effect accurately modelled by the 3D3V model is the inclination of the velocity ellipsoid, which increases with $z$. This results in an increasing contribution with $z$ of the major axis component of the ellipsoid (equal to $\sigma_R$ at $z$=0) to vertical velocities. This has the effect of increasing $\sigma_z$ with $z$ in contrast to the 1D1V model, where $\sigma_z$ remains constant for an isothermal DF.
Taking into account the tilt of the ellipsoid requires the velocity dispersions, $\sigma_z$ and $\sigma_R$, to be adjusted simultaneously.
\\

Finally, the result obtained above calls for a correction linked to the slope of the Galactic rotation curve. It is well established in the solar neighbourhood and is \,$\mathrm{d}\,V_c/\mathrm{d}\,R \sim -$1.7 km\,s$^{-1}\,$kpc$^{-1}$ \citep{2019ApJ...871..120E}, namely, $\mathrm{d}\,\log V_c / \mathrm{d}\,\log R= -$0.058. The rotation curve of the BGM used here fits the observed curves precisely \citep[see Figure 1 in ][]{2022A&A...667A..98R}, but its derivative in $R_0$ is almost zero.
If we had used the correct slope of the rotation curve, this would not change the results concerning the vertical variation of the gravitational potential. However, for the calculation of the local dark matter density, the term 
 $(1/R) \,\mathrm{d}[ R\, \mathrm{d}\Phi/\mathrm{d}R)] /\mathrm{d}R$ of the Poisson equation depends on this slope. This introduces a correction term $\rho_\mathrm{dm,corr}$ which here is $-0.0018\,$M$_\odot$pc$^{-3}$. Taking this correction into account, we finally obtain for the local dark matter density: $\rho_\mathrm{dm}(0)= 0.0146 -0.0018$= 0.0128$\pm0.0002\,$M$_\odot$pc$^{-3}$.
 \\

\begin{figure}[htbp]
   \centering
   \includegraphics[scale=0.24]{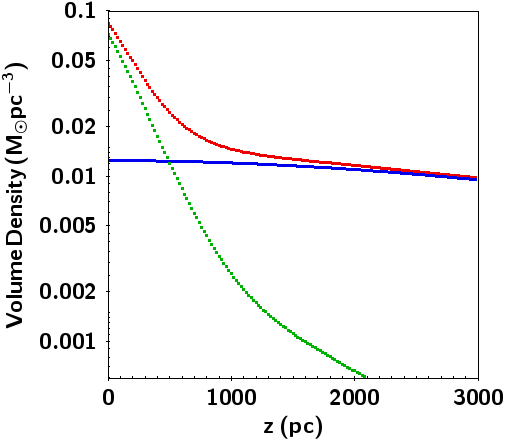} 
   \includegraphics[scale=0.24]{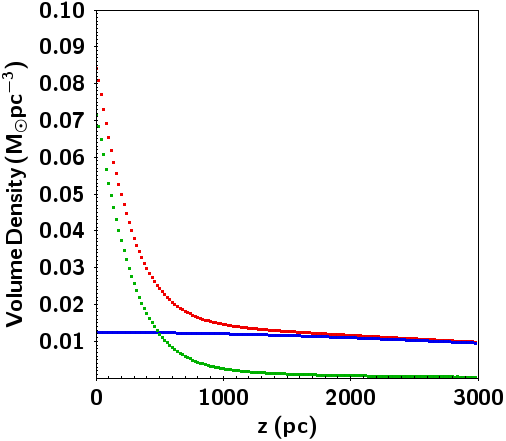} 
  \caption{Volume mass density at the solar radius position as a function of Galactic height $z$ for  baryons (green), dark matter (blue), and total (red). }
   \label{fig:f8}
\end{figure}

Thus, we get the following values for the total local density, local density of baryons and dark matter (Figure~\ref{fig:f8}): $\rho_\mathrm{tot,0}=0.0844\, $M$_\odot$pc$^{-3}$, \,
   $\rho_\mathrm{bar,0}=0.0716 \, $M$_\odot$pc$^{-3}$, \,
    and $\rho_\mathrm{dm,0}=0.0128 \, $M$_\odot$pc$^{-3}$.

\begin{figure}[htbp]
   \centering
   \includegraphics[scale=0.25]{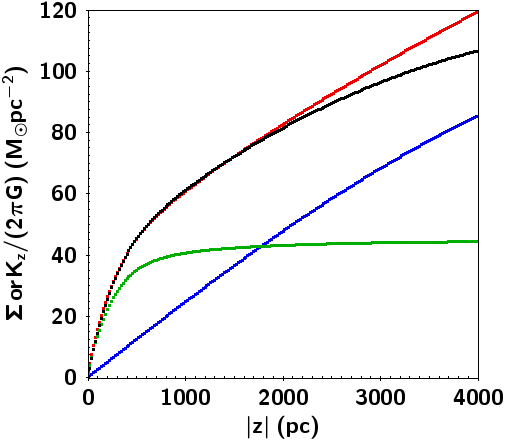} 
   \caption{Surface mass density at the solar radius position as a function of Galactic height $z$ for the baryons (green), dark matter (blue) , and  total (red). The vertical force $K_z$ is superimposed in black.
    } 
   \label{fig:f9}
\end{figure}

\noindent We obtained for the force, $K_z / (2 \pi G),  $  as seen in Figure~\ref{fig:f9}, at different heights, 0.5, 1.1, 2, and 3\,kpc, respectively: 45.4, 63.6, 81.5, and 96.4 M$_\sun$ pc$^{-2}$.
\\

\subsection{Sun velocity relative to the LSR}

 For the velocity components of the Sun relative to the LSR, we obtained $U_\odot$=12.9 km s$^{-1}$ and $W_\odot$=7.8 km s$^{-1}$,  deduced from the mean velocity of the samples.
  On the other side, the $V_\odot$ component is obtained with the best-fitting of the distribution functions. Then,
 $W_\odot$ is quite stable from $z$=500 pc to 3 kpc, while the $U_\odot$ component varies by 3\,km\,s$^{-1}$. The velocity component $V_\odot$=12 km s$^{-1}$ is derived by correcting the asymmetric drift, which changes with $z$ and by 
  adjusting $V_\odot$ so that the observed and modelled distributions are superimposed with  by posing $V_{\phi,mod}(z)$= $V_c(R_0)$+ $V_\odot$ 
 + $\overline{V_{obs}(z)}$ (Figure~\ref{fig:f7}). Then,
 $\overline{V_{obs}(z)}$ is the mean observed velocity at different height, $z$.
This is close to the values given by \citet{2021MNRAS.504..199W}  for the mean solar motion 
(11.69$\pm$0.68 , 10.16$\pm$0.51,7.67$\pm$0.10) km s$^{-1}$, 
and \citet{2010MNRAS.403.1829S}  who found (11.1$\pm$0.69, 12.24$\pm$0.47 , 7.25$\pm$0.37 ) km s$^{-1}$.
It is also comparable to the values obtained in \citet{2022A&A...667A..98R}   $(U_\odot=10.79$  km s$^{-1}$, V$_\odot$=11.06 km s$^{-1}$, and W$_\odot$=7.66  km s$^{-1}$). 
We note that our method to determine $V_\odot$ from data at large $z$ is different from these based on the extrapolation of the  asymmetric drift velocity to  $\sigma_R$ relation to null velocity dispersion. Such methods are more affected by non-stationary perturbations, whose relative amplitudes are high for small velocity dispersions and at low $z$ values.

\section{Discussion on the baryonic surface mass density}

Measuring the $K_z$ gives access to the  mass density. However, this does not allow us to distinguish between baryonic and dark matter. 
The slope of the $K_z(z)$ decreases sharply at 600\,pc when the density of baryonic matter drops rapidly and it is only around $z$=1.4\,kpc that the surface density of baryonic matter becomes less than that of the dark matter density, $\rho_\mathrm{dm}$. 
As a result, determinations of the dark matter density based on the $K_z$  measurements below $\sim1.4$\,kpc are strongly correlated with the  value of the integrated baryonic surface mass density, $\Sigma_\mathrm{bar}$. We refer to Figure 8 in \citet{2014JPhG...41f3101R} and Figure 13 in \citet{2024ApJ...960..133G}.
 Hence, an error of 10\% in the estimated surface mass density of baryons  (stars and gas) can produce a systematic error up to 35\% in the deduced density of dark matter.

More directly, \citet{1981ApJ...251...61C} determined the dark matter density by developing a global mass model of the Galaxy that includes stellar discs and an assumed spherical dark matter halo.
 The mass of the stellar discs is constrained by stellar counts  \citep[e.g.][]{1980ApJS...44...73B}.
\citet{1981ApJ...251...61C} obtained a local dark matter density of 0.011 M$_\odot $pc$^{-3}$. This was probably the first determination of the local dark matter density after \citet{1978PhDT.......195B, 1981AJ.....86.1791B,1981AJ.....86.1825B, 2023arXiv230906390B}
definitively demonstrated its existence in disc galaxies.
A simultaneous combination of the two approaches is the dynamically coherent Galactic model of global stellar population synthesis 
\citep{1986A&A...157...71R,  1987A&A...180...94B, 2022A&A...667A..98R}, which also makes it possible to constrain $\rho_\mathrm{dm}$.

In this paper, the adjustment of the counts and kinematics of the red clump giant stars is performed at a a large distance outside the plane at  $z$=3.5\,kpc, in a region where baryonic matter is almost absent. This enables the measurement of the dark matter density to be almost completely decoupled from the value of the surface density of baryonic mass. However, this is a measurement at large $z$. If we want to correctly determine,  at $z$=0, the local density of dark matter and the flattening of the dark matter halo, it is still necessary to know precisely the distribution of baryons in the disc. In this case, an uncertainty on the value of 
$\Sigma_{bar}$  also introduces another second uncertainty on the component of the dark halo. In this case, an uncertainty of 10\% on the baryonic surface density produces a systematic error of about 10\% on the local density of the dark halo. This is because the contributions at ($R_0,z$=0) of the baryons and the dark halo to the radial force, $K_R$, are of the same order of magnitude.

\paragraph{Local baryonic mass density and dark matter halo flattening:} 

To correctly evaluate the flattening of the dark halo in addition to measuring the dark matter density, it is therefore necessary to have an accurate characterisation of the distribution of the baryonic mass of the Galaxy. 
To this end, we rely here on the determinations obtained by fitting the BGM \citep{2022A&A...667A..98R} to the most recent {\it Gaia} DR3 observations of positions, parallaxes, proper motions, and radial velocities in a sphere up to about 3 kpc from the Sun.
The BGM is a  model of stellar population synthesis of the Galaxy, bringing together information on the evolutionary stages and masses of stars as a function of their ages.

\begin{figure}[htbp]
   \centering
   \includegraphics[scale=0.42]{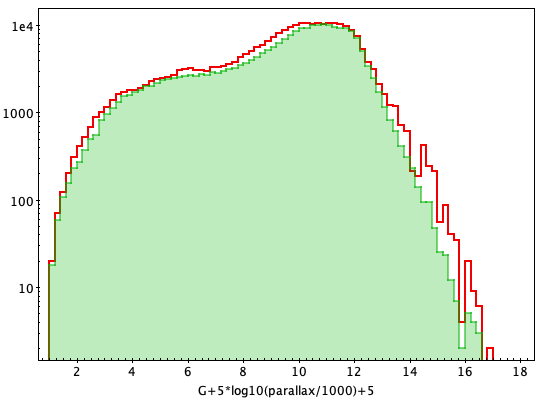} 
   \includegraphics[scale=0.42]{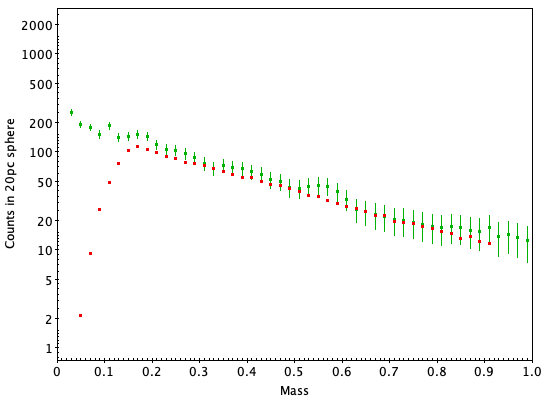} 
   \caption{ Local stellar luminosity function (top). Observations with $G$$<$17 and $\pi>$10 mas are shown in green and BGM model in red.  Present-day mass function  within 20 pc (in green), see text, and from BGM simulations (in red) with   G$<$17 (bottom).}  
   \label{fig:f10}
\end{figure}

 The luminosity function, excluding white dwarf stars,  was obtained by MCMC fitting to the local sphere 
  \cite[the {\it Gaia} catalogue of nearby stars, hereafter GCNS,][]{2021A&A...649A...6G}
 and to distant star counts in different directions \citep[see][]{2022A&A...667A..98R}. Assuming null extinction, Figure~\ref{fig:f10} presents on the top panel the distribution of absolute G magnitudes, taking into account the selection criteria (limiting magnitude G$<$17, parallaxes$>$10 mas).
The bottom panel shows the present-day mass function  for stars with mass lower than 0.9 M$_\odot$, within a sphere of 20\,pc radius centred on the Sun,
obtained by \citet{2024ApJS..271...55K} and from a BGM simulation with the limiting magnitude G$<$17 that reduces the  number of very low-mass stars.
\\
\\
This model combines  elements  rarely found together in other Galactic models. The dynamical coherence links the gravitational potential to all other Galactic components. The stellar distribution functions, density, and kinematics; hence, it is consistent with the potential.
The approach involves less uncertainty than alternative methods previously described for the following reasons: (i) the vertical and horizontal stellar density distributions are dynamically justified and (ii)  the stellar luminosity function is constrained by stellar mass luminosity relationships provided by the most recent evolutionary tracks based on {\it Gaia} observations. The Starevol stellar evolution models \citep[e.g.][]{2012A&A...543A.108L}  cover a wide mass range from 0.6 to 6 solar masses and wide ranges of metallicity and $\alpha$-abundance, which are incorporated into the BGM by \citet{2017A&A...601A..27L}. They ensure the reliability of the mass-luminosity relation in this mass range.
\\

The adjustments to this model give the local stellar surface density $\Sigma_*= 33.1\,$M$_\odot$pc$^{-2}$.
For the ISM contribution, we have assumed $\Sigma_\mathrm{ISM}= 11$\,M$_\odot$pc$^{-2}$,
which, in this case, is the least well known quantity, with an uncertainty of at least 20\%, i.e. $2\,$M$_\odot$pc$^{-2}$.
In the BGM model, the ISM density distribution is a double exponential law with scale length and height: 
$h_R$=700\,pc, $h_z$=200\,pc, and a local density 
$\rho_\mathrm{ISM,0}$=0.0275\,M$_\odot$ pc$^{-3}$.
The value of $\Sigma_\mathrm{ISM}$ adopted here is comparable to that choosen by other authors, who propose a decomposition of the ISM into sub-components. We refer, for instance, to \citet{hol06} or Table 3 in \citet{2015ApJ...814...13M}, 
but lower than in the case of \citet{2014JPhG...41f3101R}.

Thus, for all baryons, $\Sigma_\mathrm{bar}=\Sigma_{*}+\Sigma_\mathrm{ISM}=  44.1\pm 2\,$M$_\odot$pc$^{-2}$.
Our measurement with the sample of red clump giants gives for the local density of the model:

 $\rho_\mathrm{tot,0}=0.0844 \,$M$_\odot$pc$^{-3}$, \,
 
 $\rho_{*,0}$=0.0441\,M$_\odot$ pc$^{-3}$,  \,
 
  $\rho_\mathrm{ISM,0}$=0.0275\,M$_\odot$ pc$^{-3}$,    \,
  
   $\rho_\mathrm{bar,0}$=0.0716\,M$_\odot$ pc$^{-3}$,  \,
   
    and $\rho_\mathrm{dm,0}$=0.0128\,M$_\odot$pc$^{-3}$.

\noindent The stellar density can be compared for instance with the recent determination of $\rho_{*,0} = 0.043\pm0.004\,$M$_\odot$pc$^{-3} $ obtained by  \citet{2015ApJ...814...13M}.
These values are also similar to those recently published, see for example references in \citet{2024ApJ...962..165H}.
In particular, we obtain values close to 
\citet{2023MNRAS.520.1832B}, \citet{2021RPPh...84j4901D}, \citet{2015ApJ...814...13M}, \citet{2020A&A...643A..75S}, and \citet{2016ARA&A..54..529B}.

In the light of the discussions developed above, we consider that the analysis of {\it Gaia} observations using the Galactic model for the synthesis of stellar populations by \citet{2022A&A...667A..98R}  provides a more complete and rigorous approach for the determination of local densities. This is even if the version used here is a simplified, axisymmetric version of the Besan\c{c}on model.
\\

The formal random error resulting from the adjustments of the moments of the distribution functions of the stars towards the Galactic poles (Figure~\ref{fig:f7})  is very small. 
On the other hand, the uncertainty of the ISM contribution introduces the largest possible source of systematic error.
An uncertainty of 2\,M$_\odot$pc$^{-2}$ in the ISM surface density (i.e. an error of 20\%) translates into an uncertainty of 5\% in the local surface density of baryonic discs.
In order to keep the Galactic rotation curve unchanged, this also means an uncertainty of $\pm5\%$ on the mass of the dark matter halo. As we determine the dark matter density at $z$ heights  between 2 and 3.5\,kpc, this implies an uncertainty on the flattening $q_\phi$ of about 4\% and on $\rho_\mathrm{dm}$($z$=0) of 7\%.
So the flattening of the potential linked to the dark halo is $q_\mathrm{\phi,dm}=0.843\pm{0.035}$ (or a flattening of the density of the dark halo $q_\mathrm{\rho,dm} = 0.78\pm0.06$) and $\rho_\mathrm{dm}(z=0)=0.0128\pm{0.0008}$\,M$_\odot$pc$^{-3}$.

\section{Discussion}

With the arrival of the {\it Gaia} measurements, considerable progress has been made in our understanding of the Galaxy. The precision and abundance of the {\it Gaia} data have clearly revealed many properties and characteristics of the Galactic structure. 
At the same time, new, more comprehensive and detailed methods of analysis have been rapidly implemented. With regard to the local measurement of the gravity field, we now place our results in the context of recent works. 

The most original and unexpected result is certainly the new measurement of the $K_z$ based on the analysis of local non-stationary effects, in particular the spiral structures in the phase space ($z,w$) visible at $z<1$\,kpc \citep{2021A&A...653A..86W}.
 This method is totally different from the traditional approach of \citet{kap22} and \citet{oor32}, who assumed a stationary state.
 Using this method \citet{2024ApJ...960..133G}  recently obtained $\Sigma_\mathrm{tot}(z=1.1\,$kpc)=63\,M$_\odot$pc$^{-2}$, comparable to our results.
\\

In what follows we report published results of gravity field measurements obtained at large heights, $z$, for the whole Galactic disc.
\citet{2020MNRAS.494.6001N} carry out a complete modelling of the Galaxy kinematics over a range from 5 to 12 kpc in radius and up to $|z|$=2 kpc based on the older {\it Gaia} (DR2) measurements. They derive the density $\rho_\mathrm{dm}$=0.0115$\pm$0.0020\,M$_\odot$pc$^{-3}$, very close to our values. However,  they obtain a very low value of $\rho_\mathrm{tot}$=0.0640$\pm$0.0043\,M$_\odot$pc$^{-3}$ and an inaccurate measurement of the flattening of the dark halo $q=1.14\pm0.21$. 
These different results could be explained by the less precise measurements in the {\it Gaia} DR2 catalogue.

With {\it Gaia} DR3 and APOGEE data, \citet{2024ApJ...962..165H} modelled a large area of the Galaxy, with $R$ ranging from 8 to 11\,kpc and $z$ up to $\sim$ 1.5\,kpc. They got similar results to ours for the local measurements:
$\Sigma_\odot$($z$=1.1\,kpc) = 72$^{+6}_{-9}$\,M$_\odot$pc$^{-2}$,
$\rho_{tot}(z$=0) = 0.081$^{+0.015}_{-0.009}\,$M$_\odot$pc$^{-3}$,
 using the most constraining abundance ratio, [Mg/Fe]. This corresponds to
a dark matter contribution to the surface density of $\Sigma_\mathrm{dm}$($z$ = 1.1 kpc) = 24$\pm$4\,M$_\odot$pc$^{-2}$

Their approach is  innovative in the sense that they are able to split the samples into several still highly populated subsamples with various element abundances and also kinematics. These numerous subsamples allow us to put  constraints at a range of heights $z$ (unlike other works which use only two or three subsamples,  constraining the $K_z$ force at a limited number of $z$ heights).
Their second ingenious approach is that they do not fit the moments of the distribution function, density, and vertical velocity dispersion. But instead, they directly adjust the ($z,w$) distribution identifying the isolevels in this phase space, resulting in stronger constraints on the $K_z$ force. However, they applied a 1D1V dynamical model that we consider is not  appropriate to analyse the $K_z$ force at $z$ higher than 1\,kpc. They intend to improve this point in a future work.

\citet{2024MNRAS.527..959C} address the problem of measuring the Galactic gravity field for a wide range of Galactic radii and for vertical heights up to 3 kpc.
They separate the thin and thick disc populations on the basis of chemical abundances using spectroscopic measurements from the APOGEE survey.
 They obtain significantly different results for the $K_z$ with the thin or the thick disc samples. They conclude that traditional $K_z$ analyses are 'challenging' and highlight the problem of stationarity. However, if we look at their graphs, we can see that the deviations from north-south symmetry and the stationarity problems appear essentially close to the plane, with abrupt changes in the velocity dispersions below $z$=500 pc.

In particular, their measurement of the $K_z$ is close to our determination for $R$ between 8 and 9\,kpc and for their analysis of the thick disc at large $z$ up to 4 kpc (see their Figures 6 and 7, for $R$=8.5\,kpc, and their measurement of $\Sigma(z)$).
 We note, however, that their distances based on GSP-phot photometric distances deduced from {\it Gaia} BP/RP spectra and parallaxes (all from {\it Gaia} DR3) do not seem to have been corrected for the biases mentioned by \citet{2021A&A...649A...4L}.

\citet{2023MNRAS.520.1832B} also carried out a global model of the Galaxy by fitting 3D3V {\it Gaia} DR3 measurements within a neighbourhood of 3\,kpc from the Sun. The model is based on stellar distribution functions that depend on integrals of motion. 
Their dark halo is so constructed that it is dynamically consistent with the rest of the components of their Galactic model.
They fit 1D projected velocity distributions instead of moments.  As the 1D projections of the velocities are strongly non-Gaussian, this prevents valuable information from being discarded.
Their results are extremely close to what we obtained: 

$\rho_\mathrm{dm,0}$=0.0121\,M$_\odot$pc$^{-3}$,

$\Sigma_\mathrm{tot}(1.1\mathrm{kpc})$=61.2\,M$_\odot$pc$^{-2}$,

$K_z(\mathrm{1.1kpc}) /(2\pi G)$=64.1\,M$_\odot$pc$^{-2}$.

However, if we consider their Figure 13 (in \citealt{2023MNRAS.520.1832B}), the flattening of their dark matter halo seems to be close to 0.95, nearly spherical.
A provisional limitation of the model is that the adjustment of the vertical density of the disc is made using old stellar counts. While justifying this temporary choice, they have left the use of counts taken directly from {\it Gaia} observations for a future work .

\citet{2024ApJ...967...89I} has given an estimate of the halo gravitational potential from the trajectories of stellar streams. They include the  largest combination of  streams probing the halo and disc gravitational potential leading to extremely small formal errors of the gravitational forces.
This method of investigating the potential, now by far the most accurate of all, requires orders of magnitude fewer stars than conventional methods based on the Jeans equations. This translates into extreme precision, because the gravity field is determined at nearly every position of the stars within stellar streams.
These authors  found values close to ours with $\rho_\mathrm{dm,0}$=0.0114$\pm$ 0007\,M$_\odot$pc$^{-3}$
and a halo density flattening $q_\mathrm{\rho,halo}$=0.75$\pm$0.03,  similar to our value. In fact we  notice that  they set the vertical force at $z$=1.1 kpc as $K_{1.1}/(2 \pi G)$=71$\pm$6\,M$_\odot$pc$^{-2}$. This quite certainly stiffens the solution for determining the local dark matter density.
\citet{2024ApJ...967...89I} recognised  that 
tight constraints derived on the local dark matter are clearly (in part) 'an artefact of the rigidity of the analytic function'.
We note that the flattening of the dark halo they obtain is based on a global modelling of the Galactic potential, while ours is  more local.
\\

\paragraph{ Stationarity:}
All these results remain subject to the stationarity assumption, which is only partially satisfied and whose impact on the study of the $K_z$ has been studied by \citet{2014JPhG...41f3101R}. \citet{2024A&A...686A..70W} show that this impact is moderate.
They find fluctuations of only about 20\%, by tracing the local spiral arms in the Galactic plane through dynamical measurements of the gravity field close to the plane. At high $|z|$, however, the North-South symmetry is better, and the deviations between the north and south directions remain small for our samples. We have therefore ignored the disequilibria of the Milky Way, considering that these effects are weak \citep[see also the comments in][paragraph 6.4]{2024ApJ...962..165H}.
\\
\paragraph{ Other models:}
 The mild flattening of the dark  halo and the small variation of the  dark matter density between $z$=0 and 3.5 kpc do not seem to favour 
the hypothesis of late accretion of satellite galaxies, which would create a very thick disc or slowly rotating or a flattened spheroidal component of dark matter as suggested by the cosmological simulations of \citet{rea09} or \citet{2014ApJ...784..161P}; nor does this slight flattening seem to be very favourable to MOND's predictions of the presence of a thick disc of phantom matter 
\citep{2001MNRAS.326.1261M,bie09}. 
However, truly firm conclusions could only be reached by adjusting the observations of the kinematic-density profiles within the framework of an accurate modelling of the gravitational potentials predicted by these two hypotheses.

\section{Conclusion}    

In this work, we re-examine the gravity field measurements and the mass distribution in the solar neighbourhood towards the Galactic poles.
As noted by \citet{2024MNRAS.527..959C}:\ the measured surface density is 'highly dependent on the assumptions made in its calculation'.
This is certainly also true for other measured quantities such as the vertical gravitational forces or the baryonic and dark matter densities, even if the most recent works often publish values that are quite close to each other.
Here, we have carried out a study to minimise the dependence of the results on the underlying assumptions.

To this end, we measured the gravity field sufficiently far outside the Galactic plane, where the dark matter mass density largely dominates that of baryonic matter and
 where the variation of the gravity field depends almost solely on the dark matter density. 
 Thus, we have analysed the gravity field up to $z$ heights of 3.5 kpc, which is significantly higher than in previous studies.

On the other hand, for the accurate modelling of the near-plane gravity field, we assess the mass of stellar discs based directly on the most recent stellar counts and mass-luminosity relations.
This provides a more accurate measure of the surface density of baryonic matter, which is also necessary for a more robust determination of the density of dark matter within the Galactic plane. 
This point is important since it is well established that the local measurement of the dark matter density is correlated with the value adopted for the surface density of baryonic matter \citep{2014JPhG...41f3101R,  2024ApJ...960..133G}.
For that reason, our determination of the distribution of baryonic matter in the solar neighbourhood is based on recent adjustments of the stellar counts and their kinematics, the local luminosity function, and the IMF determined by \citet{2022A&A...667A..98R} and \citet{ 2017A&A...601A..27L}.

Finally, for this study, the sample of tracer stars used are red clump giants towards the Galactic poles. By limiting ourselves to bright, 'nearby' stars with distances of less than 4 kpc, we have obtained a sample with high measurement accuracy, which profoundly reduces the sources of observational bias.
The moments of the observed distributions of the stars in the sample (density, asymmetric drift and velocity dispersions $\sigma_R$, $\sigma_\phi$, and $\sigma_z$) were fitted with analytical stationary distribution functions. 
These functions depend on three integrals of motion to properly model the correlations between radial and vertical motions, which is necessary for $z$ heights greater than 1\,kpc.
\\

Our results include  
the measurement of the gravity field from $z$=0 to 3.5\,kpc, the variation of the dark matter density with $z$, and the flattening of the dark matter halo: 

$\rho_{*,0}$=0.0441 M$_\odot$pc$^{-3}$, \, 

  $\rho_\mathrm{ISM,0}$=0.0275 M$_\odot$pc$^{-3}$,  
  
   $\rho_\mathrm{bar,0}$=0.0716 M$_\odot$pc$^{-3}$, 

 $\rho_\mathrm{dm}(z=0)=0.0128\pm{0.0008}$ M$_\odot$pc$^{-3}$ = 0.486 $\pm$0.030 Gev\, cm$^{-3}$,

$q_{\phi,dm}=0.843\pm{0.0035}$ and $q_{\rho,dm}=0.781\pm{0.0055}$.
\\
We  present a model that provides a good fit to the observations.
The spatial distribution and kinematics of our stellar sample can be described 
to recover characteristics of our Galaxy.
Further progress can  be expected; for instance, instead of simply adjusting the moments of the stellar distribution functions, an adjustment of the 1D projections of the velocities would offer better constraints on the modelling.
\\
\begin{acknowledgements}
During the analysis, we have made extensive use of the astronomical java software TOPCAT and STILTS (Taylor 2005). This work has made use of data from the European Space Agency (ESA) mission {\it Gaia} (http://www.cosmos.esa.int/gaia), processed by the {\it Gaia} Data Processing and Analysis Consortium (DPAC, http://www.cosmos.esa.int/ web/gaia/dpac/consortium). This publication makes use of data products from the Two Micron All Sky Survey, which is a joint project of the University of Massachusetts and the Infrared Processing and Analysis Center/California Institute of Technology, funded by the National Aeronautics and Space Administration and the National Science Foundation.
\end{acknowledgements}

\bibliographystyle{aa} 

\end{document}